# Formation of bound states from the edge states of 2D topological insulator by macroscopic magnetic barriers

D V Khomitsky*, A A Konakov and E A Lavrukhina

Department of Physics, National Research Lobachevsky State University of Nizhny Novgorod, Nizhny Novgorod, Russian Federation

*E-mail: khomitsky@phys.unn.ru

**Abstract**

A model of bound state formation from the delocalized edge states of 2D topological insulator is derived by considering the effects of magnetic barriers attached to the edge of the HgTe/CdTe quantum well. The resulting structure has a spatial form of 1D quantum dot with variable number of bound states depending on barrier parameters. The spatial profile of exchange interaction between the edge states and barriers is derived from the interaction with single impurity magnetic moment and is generalized for the barrier bulk structure formed by ensemble of impurities. The resulting Hamiltonian is studied as a function of barrier parameters including their strength and orientation of the magnetic moments. It is shown that for parallel magnetization of two barriers at least two discrete levels are formed regardless of the barrier strength. For antiparallel magnetization at least a single bound state is formed for any strength of the barriers. Our results may help in design of novel types of quantum dots based on topological insulators.

Keywords: topological insulator, edge state, magnetic impurity, magnetic barrier, quantum dot

## 1. Introduction

Topological insulators (TI) represent a class of condensed matter structures where the Fermi level is in the bulk gap but the highly conductive edge states exist being topologically protected against the scattering on nonmagnetic impurities [1-4]. The spectrum of the edge states has a number of special points (Dirac points) near which it can be described by a linear massless dispersion relation, and the corresponding states are called the Dirac-Weyl massless fermions. During the last fifteen years the physics of TI has demonstrated a significant progress both in the description of their fundamental properties and classification as well as in the experimental design and fabrication. One of the problems that remain on the way to a wider utilization of TI in device nanostructures is the delocalized nature of the topologically protected edge states. In 3D TI the edge states are formed on the whole 2D edge surface and in 2D TI they are created on the 1D continuous or periodic edge such as in the HgTe/CdTe quantum well (QW)-based TI if the QW width exceeds certain critical value [1–5]. In contrast, in a nanodevice fabrication it is desirable to have compact objects such as quantum dots (QD) with bound states and discrete energy spectrum.

Several models of QD formed in TI has been proposed so far. They consider both 3D [6–8] and 2D [9–11] TI where the edge states are transformed from continuous to bound ones by means of different barriers. In [7] the signatures of Coulomb blockade have been traced in the conductivity patterns for the QD formed by semiconductor barriers on the surface of Bi$_2$Se$_3$ 3D TI with the typical QD size of about 100 nm. Since the pure electrostatic barrier cannot bound the massless fermions due to the Klein tunnelling [12] one can apply the barriers with magnetic structure which breaks the at least partially the protections for the edge states. The effects of the magnetic field on the TI spectrum have been studied in detail and it is known that the magnetic field opens a gap in the Dirac fermion spectrum [1, 8] and couples the counter-propagating edge states in HgTe/CdTe quantum wells [13]. A possible way to study the effects of the edge state interaction with a macroscopic magnetic barrier can be proposed from the following considerations. There is a number of models where the interaction of the edge state in TI with a single magnetic [14–16] or non-magnetic [17] impurity has been studied, as well as with an ensemble of magnetic impurities distributed in the whole bulk sample [18, 19] or at the TI / magnetic insulator interface with studies of the magnetic proximity effects [20–27], including the combination of several TI and MI layers [28]. The interaction of TI edge states with the nuclear spins via the hyperfine interaction has been considered in [29, 30] where it has been shown that an almost complete decrease in conductance at zero temperature for a long edge can be expected indicating the localization possibility by the interaction with magnetic moments. For the case of bulk distribution of magnetic impurities described in [18] it has been demonstrated that the random magnetic impurities with parallel magnetization inserted into the bulk sample can open a gap for helical edge spectrum. This gap was estimated to be of the order of the amplitude assigned to the magnetic impurity term in the Hamiltonian times the bulk impurity concentration, reaching as high as 0.5 eV for Mn-doped HgTe [18]. Within such an approach a bulk magnetic barrier with finite dimensions can be treated as an ensemble of specifically localized individual magnetic moments, so the interaction with such barrier can be treated by averaging the exchange interaction produced by a single impurity over the ensemble of the impurities. It is thus of big interest to derive a quantitative description of the bound state formation at the edge of TI produced by magnetic barriers of finite size and variable orientation and strength.

In the present paper we perform a derivation of the Hamiltonian describing the interaction of the 1D edge states in a 2D TI based on the HgTe/CdTe quantum well (QW) with the macroscopic magnetic barriers attached to the edge. The amplitude of the exchange interaction between the edge states and the magnetic impurities differs in various setups and under different estimates. It is expected to reach up to 40 meV in [31], up to 100 meV in [32], and up to 400 meV in [33]. The magnetic proximity effects causing the gap opening up to 90 meV near the Dirac point of the TI spectrum have been experimentally observed in other TI for the Bi$_2$Se$_3$ 3D TI capped with the MnSe bi-layers [34]. The experimental results on similar setups with a pair of magnetic stripes [35] or normal-superconducting junction [36] attached to the 2D TI edge have been reported recently. These results demonstrate the presence of the helical edge channel and its interference with ferromagnetic [35] or superconducting [36] regions expressed in the conductivity or resistance peculiarities. The distance between the ferromagnetic stripes in the experiment [35] has been equal to 200 μm which, according to our estimates, is too large and precludes the observation of a possible discrete level structure which could be formed between the ferromagnetic stripes. Overall, it can be stated that the full development of the experimental basis for QD formation in TI is still the matter of future research.

Our model of QD formation by the finite magnetic barriers has been introduced in a phenomenological way in the earlier papers [37–40] as a qualitative approach which requires the justification from the microscopic point of view being the subject of the present paper. Starting from the microscopic model for the interaction of the magnetic moment of a single impurity with the edge states derived in [14, 15], we arrive to the Hamiltonian describing the macroscopic magnetic barriers of different strength and orientation between which the bound states can be formed. We perform the analysis of the discrete energy spectrum and describe analytically the wavefunctions and the spin polarization. Our results uncover several important features. Namely, for the parallel magnetization of two magnetic barriers at least two discrete levels are formed for any barrier height, and the level positions are symmetric with respect to the Dirac point. As for the antiparallel barrier magnetization, a discrete level located at the Dirac point is always present regardless of the barrier height, and more levels with total number being an odd integer can be formed for higher barriers if their height exceeds an analytically defined threshold. These features point to the possibility of the QD formation in TI by the magnetic barriers with both high and low exchange interaction strength which indicates the possibility for experimental implementation of the proposed setup.

This paper is organized as follows. In Sec.2 we discuss a model of the exchange interaction between a single localized magnetic moment and the edge state which was previously derived in [14, 15], and adopt it for our sample geometry. In Sec.3 we build a model of a macroscopic magnet by considering the additive effects of the single magnetic moments discussed in Sec.2. In Sec.4 we perform the analysis of the localization length of the edge state into the neighbouring magnet with a wide energy gap based on the conventional Bernevig-Hughes-Zhang (BHZ) model [2] for TI. In Sec.5 we present some of the basic properties of the discrete spectrum, the wavefunctions and the spin polarization for the Hamiltonian of the bound states. Finally, in Sec.6 we give our conclusions.

## 2. Interaction of single magnetic impurity and the edge state

We start with the application of the known model [14, 15] for the interaction between the state in the BHZ model and a single magnetic moment of a localized impurity with the spin $\mathbf{S}=(S_x, S_y, S_z)$ localized at the point $(x_0, y_0, z_0)$:

$$V_{imp} = V(z_0) \cdot f(x-x_0, y-y_0, z-z_0), \quad (1)$$

where $f$ is a spatial profile of the impurity potential with a typical range $a$, and the spin operator $V$ has the following form in the basis of the BHZ model:

$$V(z_0) = \begin{Vmatrix} J_1 S_z & -iJ_0 S_+ & J_m S_- & 0 \\ iJ_0 S_- & J_2 S_z & 0 & 0 \\ J_m S_+ & 0 & -J_1 S_z & -iJ_0 S_- \\ 0 & 0 & iJ_0 S_+ & -J_2 S_z \end{Vmatrix}. \quad (2)$$

In (2) $S_\pm = S_x \pm i S_y$, and the real parameters $J_0, J_1, J_2, J_m$ are determined by the profile of the envelope functions $f_{1,3,4}(z_0)$ of the BHZ model [14]:

$$\begin{cases} J_0 = \dfrac{i\beta}{\sqrt{3}} f_3(z_0) \overline{f_4(z_0)}, \\ J_1 = \alpha |f_1(z_0)|^2 + \dfrac{\beta}{3}|f_4(z_0)|^2, \\ J_2 = \beta |f_3(z_0)|^2, \\ J_m = J_1 + J_0^2/J_2. \end{cases} \quad (3)$$

In (3) $\alpha$ and $\beta$ are proportional to the exchange interaction strength between the edge state and the magnetic impurity. As to the symmetry of the functions $f_{1,3,4}(z_0)$, the functions $f_{1,3}(z_0)$ are symmetric with respect to the inversion $z_0 \to -z_0$ and $f_4(z_0)$ is antisymmetric [2]. By looking onto (3) one may conclude that $J_0$ is antisymmetric and $J_m$ is symmetric with respect to the inversion. In Figure 1(a),(b) we show the spatial dependence of $J_0$ and $J_m$ as a function of the impurity position $z_0$ inside the HgTe/CdTe QW located at $[-d/2…d/2]$. In Figure 1 both parameters are scaled on the unit of coupling strength $\alpha=\beta=1$ meV and are calculated for the typical material parameters for the BHZ model [2, 5] for the QW width $d=7$ nm. From Figure 1(b) one may see that $J_m$ reaches as high as several $\alpha(\beta)$ near the QW edges with the averaged value across the QW $\langle J_m \rangle \approx \alpha(\beta)$ indicating that the values of $\alpha(\beta)$ serve as a reliable estimate for the averaged exchange interaction strength.

The basis states $\Psi_\uparrow, \Psi_\downarrow$ of the BHZ model describing localized functions which decay into the bulk have the following form [4]:

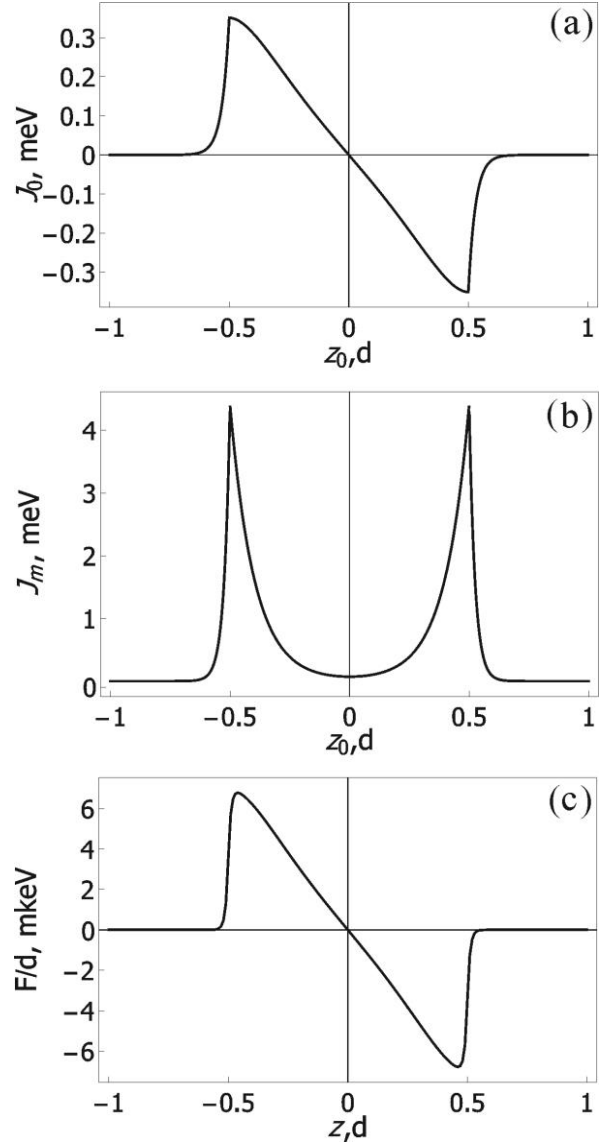

**Figure 1.** Spatial profile of the exchange interaction parameters (a) $J_0$ and (b) $J_m$ from (3) plotted as a function of the magnetic impurity position $z_0$ measured in units of the QW width $d$. (c) Antisymmetric profile of the exchange interaction (15) after integration over the individual magnets across the QW.

$$\Psi_\uparrow = \begin{pmatrix} \varphi_0(x) \\ -i\varphi_0(x) \\ 0 \\ 0 \end{pmatrix}, \quad \Psi_\downarrow = \begin{pmatrix} 0 \\ 0 \\ \varphi_0(x) \\ i\varphi_0(x) \end{pmatrix}, \quad (4)$$

where $\varphi_0(x)$ is the decaying function in the half-space $x>0$ of the bulk sample. The Hamiltonian describing the interaction of the magnetic impurity with the edge state can be written in the basis (4) by calculating the matrix elements of the operator (2) in this basis:

$$V_{\alpha\beta} = \langle \Psi_\alpha | V | \Psi_\beta \rangle, \quad \alpha,\beta=\uparrow,\downarrow. \quad (5)$$

After calculating the matrix elements (5) we arrive to the 2x2 effective Hamiltonian of the following form:

$$V(z_0) = \begin{Vmatrix} (J_1+J_2)S_z - 2J_0 S_x & J_m(S_x - iS_y) \\ J_m(S_x + iS_y) & -(J_1+J_2)S_z + 2J_0 S_x \end{Vmatrix}, \quad (6)$$

which can be composed as a combination of the 2x2 Pauli matrices:

$$V(z_0) = \left((J_1+J_2)S_z - 2J_0 S_x\right)\sigma_z + J_m\left(S_x \sigma_x + S_y \sigma_y\right) \quad (7)$$

The Hamiltonian of the form (7) has been obtained previously in [15]. Below we discuss the contributions to the exchange operator resulting after averaging the expression (7) over the macroscopic magnet comprising the individual magnetic moments with symmetrical spatial distribution centered at the QW plane of symmetry $z=0$.

## 3. Interaction with macroscopic magnet

### *3.1 Three-dimensional form of the interaction operator*

The expression (7) describes the spin part of the interaction of the edge state with a single magnetic impurity attached to a fixed point $(x_0, y_0, z_0)\equiv(x_n, y_n, z_n)$. If this individual magnet is described by the magnetization $\boldsymbol{\mu}_n$ and the g-factor $g_n$ then the exchange interaction of the edge state spin $\mathbf{s}$ with the ensemble of magnets can be described by the operator (in units where $\hbar=1$)

$$H_{exch} = -\frac{1}{2}\mathbf{J}(x,y)\cdot\mathbf{s}, \quad (8)$$

where the vector operator

$$\mathbf{J} = \frac{1}{\mu_B}\sum_n \boldsymbol{\mu}_n \frac{2}{g_n} V(z_n) J(\mathbf{r}-\mathbf{r}_n). \quad (9)$$

In (9) $V(z_n)$ is the spin part of the exchange operator from (7) and $J(\mathbf{r}-\mathbf{r}_n)$ is the exchange integral between the edge state and the single magnet. Below we will calculate the explicit form of (9) under the following assumptions. First, let us assume that all of the magnetic moments comprising the macroscopic magnet are aligned in the QW plane ($xy$) and are parallel so their $x$- and $y$- projections of spin $S$ can be characterized by a single angle $\theta$:

$$\begin{cases} S_x = S\cos\theta \\ S_y = S\sin\theta \\ S_z = 0 \end{cases} \quad (10)$$

With $S_z=0$ the term in (7) proportional to $S_z$ vanishes. The values of g-factor will also be considered as equal for all the magnetic moments. Second, the number of magnetic moments in the macroscopic magnet is large enough for converting the sum in (9) into the integral, $\sum_n \to \int \rho\, dx\, dy\, dz$, where $\rho$ is the spatial density of the magnetic moments which for the magnet with $N$ individual magnetic moments in the parallelepiped with dimensions $L_x$, $L_y$, $L_z$ can be calculated as $\rho = N/(L_x L_y L_z)$. Third, we assume a simple approximation for the exchange integral $J(\mathbf{r}-\mathbf{r}_n)\equiv J(x-x_n, y-y_n, z-z_n)$ in an exponential form with the typical range $a$ and amplitude $I$,

$$J(\mathbf{r}-\mathbf{r}_n) = I\exp\left(-\frac{|\mathbf{r}-\mathbf{r}_n|}{a}\right). \quad (11)$$

We consider a simple model for the macroscopic magnet which has a form of a parallelepiped located symmetrically with respect to the QW plane $z=0$. For such magnet geometry we may perform a transformation of the exponent argument in (11) which allows splitting of the variables. Namely, we replace the Euclidian distance

$$\frac{|\mathbf{r}-\mathbf{r}_n|}{a} = \frac{\sqrt{(x-x_n)^2+(y-y_n)^2+(z-z_n)^2}}{a}, \quad (12)$$

by the expressed with a corrected scale $a_1$ in the following form allowing to split the spatial variables in (12):

$$\frac{|\mathbf{r}-\mathbf{r}_n|}{a_1} = \frac{|x-x_n|+|y-y_n|+|z-z_n|}{a_1}. \quad (13)$$

A direct analysis of (12) and (13) indicates that for $a_1=1.5\,a$ the two expressions yield very close results in a wide limits of about [-100 $a$, …, 100 $a$] which covers the required range of exchange interactions.

Under the assumptions mentioned above the operator (9) can be written in the following form:

$$\mathbf{J} = \boldsymbol{\mu}\frac{2\rho I}{\mu_B g}\int V(z_n) \times f(x-x_n)f(y-y_n)f(z-z_n)\,dx_n\,dy_n\,dz_n \quad (14)$$

where $f(z-z_n)=\exp(-|z-z_n|/a_1)$ and similar for the other coordinates. The approximation (14) allows us to divide the averaging over the magnet into the averaging across the QW direction parallel to the $z$ axis and the averaging in the QW plane ($xy$) since the exchange parameters (3) and $V(z_n)$ in (14) depend only on coordinate $z_0=z_n$.

### 3.2 Averaging across the QW

Our averaging scheme is the following: we firstly integrate over the distribution of the individual magnets at points $z_n$ in the QW and then we average the obtained result over the QW along the $z$ coordinate. The last averaging reflects transition from 3D to 2D picture in description of the edge states. Under the continuum approximation made above the exchange interaction profile $F(z)$ across the QW direction $z$ is described by the integral over the magnets ensemble taken with the function $\exp(-|z-z_n|/a_1)$ describing the individual interaction profile times the factor $J_0(z_n)$ from (3) defining the impact of an individual magnet at the coordinate $z_n$. The distribution of the magnetic moments is considered to be symmetric and uniform along the $z_n$ inside the QW with thickness $d$. First of all, we are interested in the averaging of the term proportional to $J_0$ in the operator (7). From (14) one obtains that

$$F(z) = \int_{-d/2}^{d/2} J_0(z_n)\, e^{-\frac{|z-z_n|}{a_1}}\, dz_n \quad (15)$$

We calculate (15) by using the antisymmetric form of $J_0(z_n)$ shown in Figure 1(a). In Figure 1(c) we plot the function (15) for $\beta=1$ meV and for the ratio $a_1/d=0.01$ which seems to be realistic for the given QW width $d=7$ nm and a small exchange coupling radius although the symmetry of $F(z)$ does not depend on the specific values of the parameters. One can see that $F(z)$ is antisymmetric, and its average value across the QW vanishes,

$$\int_{-d/2}^{d/2} F(z)\,dz = 0 \quad (16)$$

By combining (16) with the chosen individual magnet orientation (10) with $S_z=0$ one may conclude that the first term in (7) proportional to $\sigma_z$ vanishes after averaging across the QW.

Let us turn our attention to the second term in (7) proportional to $J_m$. Since $J_m$ is symmetrical with respect to the QW plane of symmetry $z=0$ as it is seen in Figure 1(b) it has a nonzero average value

$$\frac{1}{d^2}\int_{-d/2}^{d/2}\int_{-d/2}^{d/2} J_m(z_n)\, e^{-\frac{|z-z_n|}{a_1}}\, dz_n\, dz = \langle J_m \rangle. \quad (17)$$

By taking into account (10), (14), (16) and (17) we arrive to the following form of the operator (7) which coincides with the one used in our previous papers [37–40]:

$$V_m = M(\sigma_x \cos\theta + \sigma_y \sin\theta)F(x,y). \quad (18)$$

where

$$\begin{cases} F(x) = \dfrac{1}{a_1}\int e^{-\frac{|x-x_n|}{a_1}}\, dx_n, \\ F(y) = \dfrac{1}{a_1}\int e^{-\frac{|y-y_n|}{a_1}}\, dy_n. \end{cases} \quad (19)$$

In (18) $M = \langle J_m \rangle S$ is the amplitude of the exchange interaction with the magnets averaged across the QW and expressed in units of energy.

### 3.3 Averaging in the QW plane

We now proceed with evaluating the explicit form of the function $F$ from (19) determining the spatial magnetization profile along the coordinates $x$ and $y$ in the QW plane. Under the approximation (13), (14) it has the form of an integral

$$F(x) = \frac{1}{a_1}\int_{x_1}^{x_2}\exp\left(-\frac{|x-x_n|}{a_1}\right)dx_n. \quad (20)$$

The integration limits in (20) are determined by the magnet spatial dimensions $L_x=x_2-x_1$ and similar for the other coordinate. The integral (20) is of the elementary type, giving us the following form of the function $F$ in different spatial areas:

In the area of the magnet for $x_1 < x < x_2$

$$F_{in}(x) = \left(2 - \exp\left(-\frac{x - x_1}{a_1}\right) - \exp\left(-\frac{x_2 - x}{a_1}\right)\right), \quad (21)$$

in the area to the left of the magnet for $x < x_1$

$$F_{left}(x) = \left(\exp\left(-\frac{x_1 - x}{a_1}\right) - \exp\left(-\frac{x_2 - x}{a_1}\right)\right), \quad (22)$$

and in the area to the right of the magnet for $x > x_2$

$$F_{right}(x) = \left(\exp\left(-\frac{x - x_2}{a_1}\right) - \exp\left(-\frac{x - x_1}{a_1}\right)\right). \quad (23)$$

We assume that the macroscopic magnet size typical exceeds the exchange coupling distance $a_1$ by approximately two orders of magnitude. In Figure 2 we show the profile described by the functions (21)–(23) for the parameters $x_1$=0, $x_2$=100 nm, and $a_1$=5 nm, plotted in dimensionless units. The high value of $a_1$ is taken for better visualization of the spatial profile of $F(x)$. The magnet borders are shown as vertical lines. One can see that the profile in Figure 2 has a step-like form which coincides with the one used in our previous papers [37–40]. It indicates a simple property of a short-range exchange interaction being active only inside the magnet area and on short distances ~ $2a_1$ near it. The exchange interaction range $2a_1$ should be compared with the penetration length $l$ of the edge state decaying into the neighboring magnet. We will estimate this length in the following section.

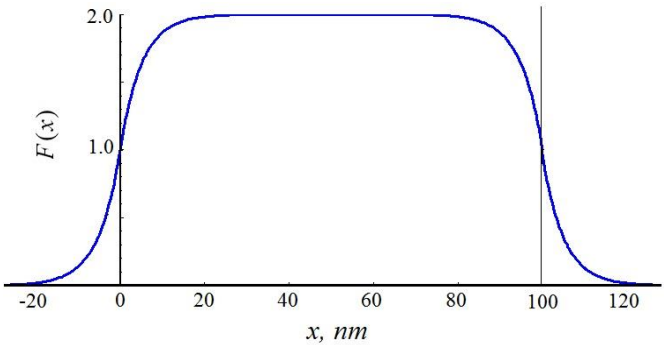

**Figure 2.** Spatial profile of the exchange interaction (21)–(23) generated by a macroscopic magnet located between $x_1$=0 and $x_2$=100 nm with the high effective exchange interaction radius $a_1$=5 nm taken for better visualization.

As to the amplitude $V_0$ of the exchange interaction, the expression (14) indicates that it differs from the amplitude of the exchange interaction with a single magnet by a factor $N_a = \rho(2a_1)^3$ where the density of the magnetic moments has been introduced during the transition from (9) to (14). The numerical value of $N_a$ corresponds to the number of magnetic moments in a cube with the size $2a_1$ where $a_1$ has the same order as the exchange interaction range $a$. Since typically it exceeds the lattice constant by a factor of 3…5, the value of $N_a > 1$ which is greater than $N_0$=1 being the value for a single magnet. As a result, the value of the exchange integral in (14) is greater than for a single magnet and can reach the amplitudes in the range of 5…40 meV in accordance with previously published estimates [31–33].

## 4. Localization length in BHZ model with a gap

Here we will estimate the penetration depth $l$ of the edge state into the space of an adjacent dielectric magnet. We take into account the influence of the magnet by introducing an additional gap $V_0$ into the BHZ Hamiltonian $H_0$:

$$H = H_0 + V_0 \sigma_z. \quad (24)$$

In (24) the parameter $V_0 > 0$ which corresponds to a trivial insulator in the area of the magnet. The value of $V_0$ is large compared to the typical edge state energy $E$ and the BHZ band width $|M|$.

Following [4], we will write down and solve the stationary Schrödinger equation for the BHZ model with Hamiltonian (24) for the edge states described by a two-component wavefunction $\psi = (\psi_1, \psi_2)$. For the quantum number $k_y$=0 where $k_y$ corresponds to the motion along the edge described by the coordinate $x$=0 the Schrödinger equation has the following form:

$$\left[\tilde{\varepsilon}(-id/dx)\hat{1} + \begin{pmatrix} \tilde{M}(-id/dx) & -iAd/dx \\ -iAd/dx & -\tilde{M}(-id/dx) \end{pmatrix} + V_0\sigma_z\right]\begin{pmatrix}\psi_1\\\psi_2\end{pmatrix} = \quad (25)$$
$$= E\begin{pmatrix}\psi_1\\\psi_2\end{pmatrix}$$

where $\tilde{\varepsilon}(k_x) = C - Dk_x^2$ and $\tilde{M}(k_x) = M - Bk_x^2$ with the operator $k_x^2 = -d^2/dx^2$. The values of the parameters $A,B,C,D,M$ can be taken from the well-known tables for TI with the specific width of the HgTe/CdTe quantum well [1–5]. We will build the solution for equation (25) in the form of a two-component spinor with exponential dependence on the $x$ coordinate decaying into the adjacent magnet area $x$<0 on a typical inverse localization length $\lambda$:

$$\begin{pmatrix}\psi_1\\\psi_2\end{pmatrix} = \begin{pmatrix}\alpha\\\beta\end{pmatrix}\exp(\lambda x). \quad (26)$$

After substituting (24) into (23) we obtain a system of linear algebraic equations for $\alpha, \beta$:

$$\left\| \begin{array}{cc} C+M+V_0+\lambda^2(D+B)-E & -iA\lambda \\ -iA\lambda & C-M-V_0+\lambda^2(D-B)-E \end{array} \right\| \begin{pmatrix} \alpha \\ \beta \end{pmatrix} = \begin{pmatrix} 0 \\ 0 \end{pmatrix}. \quad (27)$$

A nontrivial solution for the system (27) exists if its determinant is equal to zero:

$$\Delta(\lambda, E, V_0) = 0. \quad (28)$$

Equation (28) determines the inverse localization length $\lambda$ of the edge state into the adjacent magnet area as a function of the edge state energy $E$ and the magnet band gap $V_0$. After calculating the determinant in (28) we obtain and solve a bi-quadratic equation for $\lambda$ and plot the dependence of the localization length $l_{1,2}$

$$l_{1,2}(E, V_0) = \frac{1}{\lambda_{1,2}(E, V_0)} \quad (29)$$

as a function of two parameters $(E, V_0)$. In Figure 3 the dependence of $l_{1,2}$ from (29) on $(E, V_0)$ is shown for typical values of the parameters for the HgTe/CdTe QW with the width $d=7$ nm taken from [4]: $A=3.65$ eV·Å, $B=-68.6$ eV·Å$^2$, $C=0$, $D=-51.2$ eV·Å$^2$, $M=-0.01$ eV. In Figure 3 the limits of the parameters $(E, V_0)$ are $-0.075$ eV $< E < 0.075$ eV and $0.1 < V_0 < 1.0$ eV which correspond to typical values of the energy of the edge state and the band gap of a magnetic dielectric material. We focus on the largest of two values being $l_1$ shown in Figure 3a which describes the scale for the overlap of the edge state and the wavefunctions of individual magnetic impurities which form the macroscopic magnet. Looking onto Figure 3a one can conclude that for typical case when $V_0 > |E|$ this localization length weakly depends on the edge state energy and decreases from $l_1 \sim 4$ nm at $V_0 \sim 0.2$ eV to $l_2 \sim 2$ nm at $V_0 > 0.4$ eV, and slowly varies for higher values of $V_0$. Such values of the penetration depth into the magnet are in agreement of the other studies of the exchange interaction models with the same values of the exchange integral [31–33].

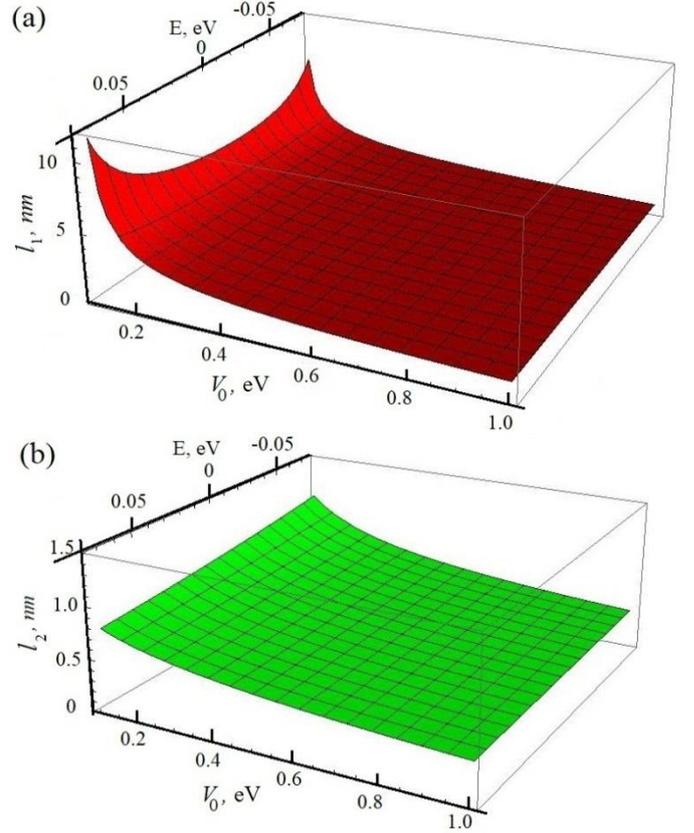

**Figure 3.** Dependence of the localization length (29) on the edge state energy $E$ and the band gap $V_0$ of the magnetic dielectric forming the magnetic barrier, shown for the (a) maximal length $l_1$ and (b) minimal length $l_2$.

## 5. Bound states between the barriers

### 5.1 Energy levels and wavefunctions

In this Section we consider the discrete states formed at the edge of TI with the presence of two magnetic barriers each of them having the spatial profile described in Section 3 by Eq.(21)–(23) and shown in Figure 2. If two semi-infinite (or, equivalently, very wide) barriers are attached to the TI edge with the spacing $L$ between them then the edge states can be described by the Hamiltonian

$$H = Ak_y\sigma_z - M_1 S(-y)(\sigma_x \cos\theta_1 + \sigma_y \sin\theta_1) \\ - M_2 S(y-L)(\sigma_x \cos\theta_2 + \sigma_y \sin\theta_2). \quad (30)$$

The Hamiltonian (30) has been introduced in a phenomenological way in our papers [35–38]. The first term in (30) is the Hamiltonian of the 1D edge states with the parameter $A=360$ meV·nm for typical TI based on HgTe/CdTe QW [1–5]. Two other terms describe the interaction with magnetic barriers discussed in Sec. 2-4. We use the step function as an approximation for the function $S(y)$ in (30) which replaces the expressions (21)–(23) with good accuracy for wide barriers. The parameter $L$ is the spacing between them being the size of the quantum dot actually formed, $\theta_1$ and $\theta_2$ are the angles of magnetization in the $(xy)$ plane for each of the barriers, and $M_1$ и $M_2$ are the exchange interaction amplitudes representing the height of each barrier. The layout of our structure is shown schematically in Figure 4. The barriers are plotted with a moderate width only for visual convenience. The discrete states of the Hamiltonian (30) formed between the barriers shown in Figure 4 can be described in the following form depending on the location in the QD area ($0<y<L$), in the left barrier ($y<0$), and in the right barrier ($y>L$):

$$\psi_{QD} = \begin{pmatrix} C_1 \exp(iEy/A) \\ C_2 \exp(-iEy/A) \end{pmatrix}, \quad (31)$$

$$\psi_{y<0} = B \begin{pmatrix} 1 \\ \alpha_1 \end{pmatrix} \exp(\gamma_1 y), \quad (32)$$

$$\psi_{y>L} = D \begin{pmatrix} 1 \\ \alpha_2 \end{pmatrix} \exp(-\gamma_2 y), \quad (33)$$

where the following energy-dependent parameters are introduced:

$$\begin{cases} \alpha_1 = -\dfrac{i\sqrt{M_1^2 - E^2} + E}{M_1} e^{i\theta_1}, \\ \alpha_2 = \dfrac{i\sqrt{M_2^2 - E^2} - E}{M_2} e^{i\theta_2}, \\ \gamma_{1,2} = \dfrac{\sqrt{M_{1,2}^2 - E^2}}{A}, \end{cases} \quad (34)$$

where $|\alpha_{1,2}|=1$. The coefficients $C_{1,2}$, $B$, $D$ in (31)–(33) are determined from the boundary conditions describing the continuity of the wavefunction at the borders $y=0$ and $y=L$ between the QD and the magnetic barriers:

$$\begin{cases} \psi_{y<0}(0) = \psi_{QD}(0), \\ \psi_{QD}(L) = \psi_{y>L}(L). \end{cases} \quad (35)$$

It should be noted that we do not address here the possible effects of the barrier finite width which is supposed to be so high that the wavefunctions (32), (33) decay almost completely before reaching the other barrier wall. That is why we neglect these far walls in the present model. The effects of the finite barrier width lead to the finite barrier transparency. This, in turn, means the formation of the quasistationary states between the barriers with finite lifetimes which were calculated in [40] for the Hamiltonian of the same type as in Eq. (30).

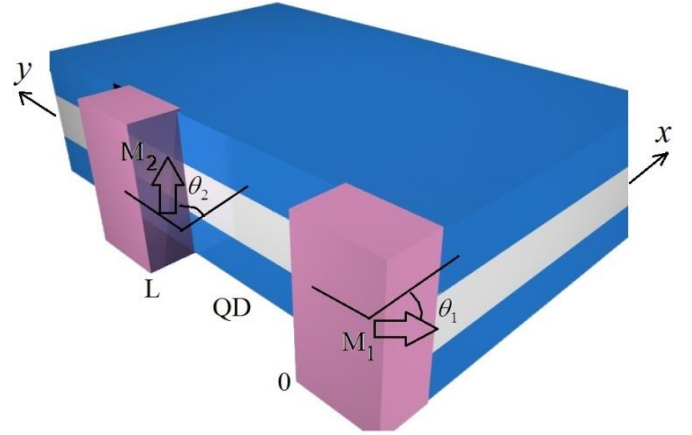

**Figure 4.** Principal scheme of the quantum dot formed at the 1D edge of the topological insulator by application of two magnetic barriers and described by the Hamiltonian (30). $L$ is the spacing between them being the size of the quantum dot (QD) formed between the barriers, $\theta_1$ and $\theta_2$ are the angles of magnetization in the $(xy)$ plane for each of the barriers, and $M_1$ и $M_2$ are the exchange interaction amplitudes representing the height of each barrier.

After substituting Eq.(31)–(33) into the boundary conditions (35) one obtains the system of linear homogeneous equations for the coefficients $C_{1,2}$, $B$, $D$. The existence of nontrivial solution is maintained by the zero determinant of this system leading to the following equation for the energy $E$:

$$\left(E + i\sqrt{M_1^2 - E^2}\right)\left(E + i\sqrt{M_2^2 - E^2}\right)/(M_1 M_2) = \\ = \exp(2iEL/A + i(\theta_2 - \theta_1)). \quad (36)$$

For the infinitively high barriers the equation (36) can be solved analytically [9], but for a more realistic situation with finite values of the barrier height $M_1$ и $M_2$ only the numerical solution is possible for a general case. After obtaining the energy $E$ of a discrete level one may normalize the wavefunction (31)–(33) to unity in the usual way: after setting, for example, $C_1=1$, one obtains the normalized eigenstate as

$$\psi_{QD} = \frac{1}{\sqrt{G}} \begin{pmatrix} \exp(iEy/A) \\ \alpha_1 \exp(-iEy/A) \end{pmatrix}, \quad (37)$$

$$\psi_{y<0} = \frac{1}{\sqrt{G}} \begin{pmatrix} 1 \\ \alpha_1 \end{pmatrix} \exp(\gamma_1 y), \quad (38)$$

$$\psi_{y>L} = \frac{1}{\sqrt{G}} \begin{pmatrix} 1 \\ \alpha_2 \end{pmatrix} \exp(-\gamma_2(y-L) + iEL/A), \quad (39)$$

where $\alpha_{1,2}$ and $\gamma_{1,2}$ are determined in (34) and the normalization constant $G$ is given by

$$G = 2L + \frac{1}{\gamma_1} + \frac{1}{\gamma_2}. \quad (40)$$

### 5.2 Existence of discrete levels in the quantum dot

Eq.(36) has several properties regarding its solutions, the main of them read as follows:
1. Any solution $E$ for Eq.(36) is a real number since the left hand side of (36) is a complex number $z$ with $|z|=1$, and the right hand side has the form of $\exp(i\varphi)$ meaning that the phase $\varphi = 2EL/A + \theta_2 - \theta_1$ is a real number;
2. The spectrum of the discrete states is bound by the lowest of the magnetic barriers height such as $|E| < \min(M_1, M_2)$;
3. The levels depend only on the difference $\theta_2-\theta_1$ between the magnetization angles;
4. If $\theta_2-\theta_1=\pi$, i.e. the barrier magnetizations are antiparallel, then for any barrier height $M_{1,2}$ a special level with the energy $E=0$ exists in the QD. This level has a number of interesting properties which will be discussed later. In particular, for this level the inverse localization length $\gamma_{1,2}$ for the state in the barrier defined in (34) reaches its maximal value which corresponds to the maximally localized state between the barriers.

As to the general rule defining the existence of solutions for Eq.(36), its depends strongly on the orientation of the barrier magnetization. If the two barrier magnetizations are equal in magnitude $M_{1,2}=M$ and their magnetizations are parallel, say, $\theta_1= \theta_2=0$, then for any value of the barrier height two discrete levels exist in the QD. This can be seen after rewriting Eq.(36) in the following form:

$$f_1(E) = f_2(E) \quad (41)$$

where

$$\begin{cases} f_1(E) = \text{ArcTan}\left[\dfrac{E\left(\sqrt{M_1^2-E^2}+\sqrt{M_2^2-E^2}\right)}{E^2-\sqrt{(M_1^2-E^2)(M_2^2-E^2)}}\right], \\ f_2(E) \,(\text{mod } 2\pi) = \dfrac{2EL}{A} + (\theta_2-\theta_1). \end{cases} \quad (42)$$

Let us plot the functions in the left and right side for Eq.(41) with the help of Eq.(42) for a specific value of the barrier height, say, $M_{1,2}=20$ meV, and for typical value of the QD width $L=40$ nm with the results shown in Figure 5a. By analyzing the functions (42) in Figure 5a one can conclude that two energy levels exist for any barrier height $M$ in this configurations since the plots for $f_{1,2}(E)$ always have two crossing points. This finding is confirmed by numerical solution of Eq.(36) or Eq.(41). In Figure 6a we plot the dependence of two discrete level positions on the barrier height $M$ for two equal barriers with $M_{1,2}=M=20$ meV and for parallel barrier magnetization angles $\theta_1=\theta_2=0$. From Figure 6a it follows that even for small values of the amplitude of the exchange interaction (18) the formation of two bound states is possible. The level spacing is constrained by the condition $|E| < \min(M_1, M_2)$, but very large values of $M$ are not profitable since even for large variations of the barrier height the two levels are confined within the range $|E|<10$ meV as it can be seen from Figure 6a. In Figure 6b we show the dependence on $M$ of the inverse penetration length $\gamma$ of the bound states into the barriers defined in (34).

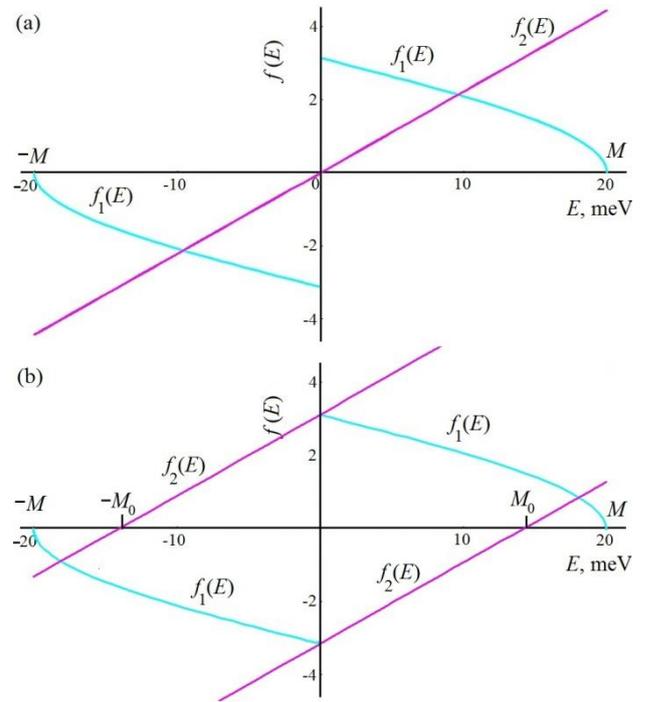

**Figure 5.** Solutions for Eq.(41) with functions $f_{1,2}(E)$ defined in (42) shown for QD width $L=40$ nm. (a) Solutions for the equal barrier height $M_{1,2}=M=20$ meV and for parallel barrier magnetization angles $\theta_1= \theta_2=0$. Two energy levels exist for any barrier height $M$ in this configurations since the plots for $f_{1,2}(E)$ always have two crossing points. (b) Solutions for the antiparallel barriers with $\theta_2-\theta_1=\pi$. The solutions with nonzero energy exist only when $M > M_0$ where $M_0$ is defined in (43).

One can see that at low barrier height (that is, for low exchange coupling between the edge states and the magnets) the bound states are poorly localized since $1/\gamma$ is much lower than the QD size $L=40$ nm. As the barrier height grows, the localization of the bound states becomes more pronounced. The penetration length $1/\gamma$ reaches 50…20 nm for M=12…20 meV indicating that the thick enough barriers with the size along the TI edge being larger than 50…100 nm can be treated as semi-infinite as it is considered in our model.

Another result for the energy levels can be obtained by considering their dependence on the relative orientation of two barrier magnetizations. Let us choose the typical values of the QD width and equal barrier height: $L$=40 nm and $M_1$=$M_2$=20 meV. After solving Eq.(36) one may plot the energy level dependence on the magnetization angle $\theta_2$ for the second barrier if the magnetization at the first barrier is fixed, for example, at $\theta_1$=0. The results are shown in Figure 7.

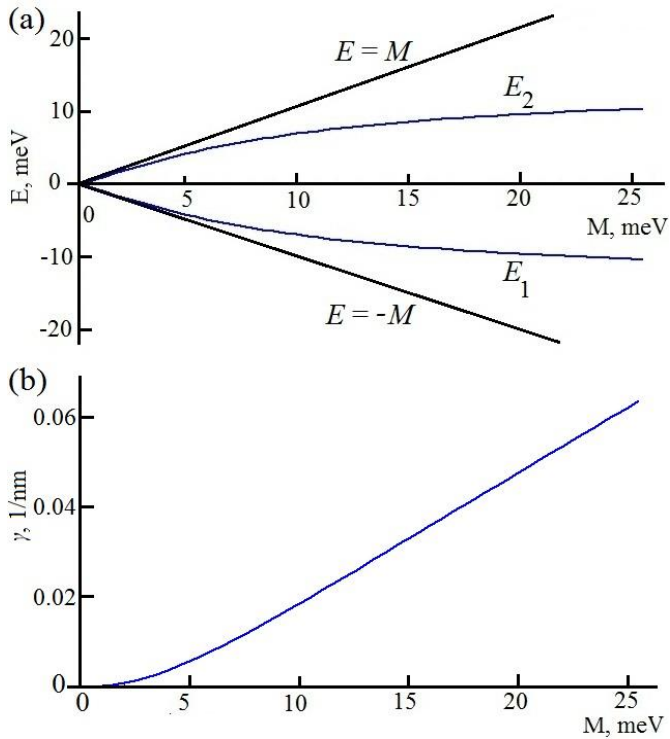

**Figure 6.** (a) Position of two discrete levels $E_{1,2}$ for parallel barrier magnetization in the interval of equal barrier height $0<M<25$ meV for QD width $L$=40 nm shown as a function of the barrier height $M$ proportional to the exchange coupling between the edge states and the magnets. Angled lines mark the limits of the discrete spectrum $|E|=M$. (b) Dependence on $M$ of the inverse penetration length $\gamma$ of the bound states into the barriers defined in (34).

Here one may see that for equal magnetic barriers with parallel magnetization $\theta_2=\theta_1=0$ two discrete levels are formed with energies $E_{1,2}=\pm 9.62$ meV being symmetrical with respect to the Dirac point $E$=0. When the orientation angle $\theta_2$ of the second barrier begins to differ from zero the levels are shifted, and approximately at $\theta_2$=$3\pi/4$ the third discrete level appears from the top of the energy spectrum. At $\theta_2-\theta_1=\pi$ when the barriers are antiparallel one has three discrete levels formed in the QD. Two of them have symmetrical energies $E_{1,3}=\pm 18.5$ meV for the present values of the barrier height while the central level always has the energy $E_2$=0.

If the barriers have non-parallel magnetizations and/or unequal height, the number of discrete levels available in the quantum dot may vary significantly. As we have mentioned earlier, for the antiparallel magnetizations $\theta_2-\theta_1=\pi$ the level $E_0$ exists for any barrier height. As to the other levels, their existence depends on the barrier height. Let us consider two equal barrier heights $M_{1,2}=M$ and antiparallel magnetizations $\theta_1$=0, $\theta_2$=$\pi$. The analysis of Eq.(41), (42) analogous to the one presented in Figure 5a leads to the following conclusion: the plots for the functions (42) have crossing points other than $E_0$=0 leading to the formation of new levels if the barrier height exceeds the critical value, i.e. for

$$M > M_0, \quad M_0 = \frac{\pi A}{2L}. \quad (43)$$

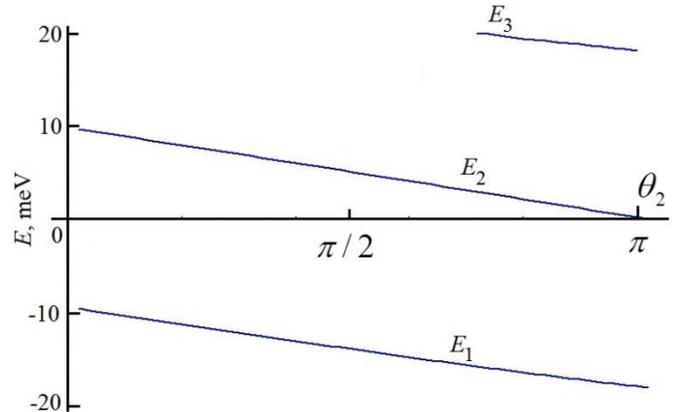

**Figure 7.** Energy level dependence from Eq.(36) with $L$=40 nm and $M_1$=$M_2$=20 meV shown as a function of the magnetization angle $\theta_2$ for the second barrier if the magnetization at the first barrier is fixed at $\theta_1$=0.

The derivation of the condition (43) is straightforward: if we substitute $\theta_2-\theta_1=\pi$ into (42) and plot the function $f_2(E)$ modulo $2\pi$ then we will see that it crosses the $f$=0 line ($Ox$ axis) at points $E=-M_0$ and $E=M_0$ where $M_0$ is given in (43), as it can be seen in Figure 5b. The left (right) part of the plot of the first function $f_1(E)$ has the leftmost (rightmost) point at $E=-M$ ($E=M$) on the same line $f$=0 where $M$ is the height of both barriers. In order for the level $E$ with nonzero energy to exist in addition to the level $E_0=0$ in accordance with (42) these lines defined by $f_1(E)$ and $f_2(E)$ need to cross as it happens in Figure 5b. Such crossing may take place only when the plot for $f_1(E)$ has the leftmost (rightmost) point at $E=\pm M$ to be located further from the coordinate origin than the one for $f_2(E)$ which means that the condition $M>M_0$ should be satisfied giving us Eq.(43). For the present set of parameters, one gets from (43) that $M_0=14.13$ meV. If $M>M_0$, one may expect the formation of three discrete levels $E_0$, $E_1$, $E_2$ in the QD with antiparallel barriers of equal height.

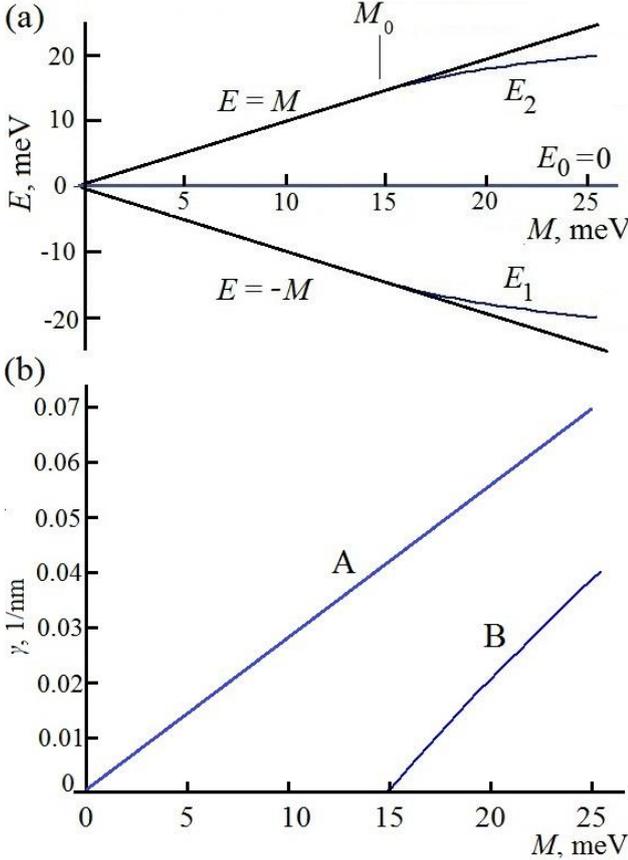

**Figure 8.** (a) Position of three discrete levels in the QD formed by barriers with equal height $M$ but with antiparallel magnetization $\theta_1=0$, $\theta_2=\pi$, shown as a function of $M$. The central level $E_0=0$ exists for any barrier height while two other levels appear when the condition $M>M_0$ is fulfilled where $M_0=14.13$ meV is defined in (43). (b) Dependence on $M$ of the inverse penetration length $\gamma$ of the bound states into the barriers defined in (34) shown for the central level $E_0=0$ (line A) and for the upper and lower levels $E_{1,2}$ (line B).

This is confirmed in Figure 8a where the level position is plotted as a function of the barrier height $M$ together with the limiting lines $|E|=M$. The numerically obtained plots in Figure 8a confirm the analytical result (43) for the existence of two levels $E_{1,2}$ with nonzero energy. In Figure 8b the dependence on $M$ is shown for the inverse penetration length $\gamma$ of the bound states into the barriers defined in (34). Line A is for the central level $E_0=0$ and line B is for the upper and lower levels $E_{1,2}$ which emerge under the condition $M>M_0$. We may see that the bound state at the central level $E_0=0$ has a higher degree of localization which reflects the general trend for a bound state of being stronger localized as soon as its energy moves away from the barrier top.

*5.3 Charge and spin density*

The expressions (37)–(39) for the wavefunctions allow the analytical calculations of the spatial distributions for the charge density $|\psi(y)|^2$ and the spin density components (in units of $\hbar/2$)

$$S_k(y) = \psi^+\sigma_k\psi, \qquad k=x,y,z. \qquad (44)$$

From (37)–(39) and (44) we obtain that the $z$-component of the spin density is identically zero, i.e. the spins of the discrete states in the QD are oriented in the ($xy$) plane just like the magnetization of the barriers. In the QD area $0<y<L$ the $x$- and $y$-components of the spin density are

$$S_{xQD}(y) = \frac{2}{G}\left(-\frac{E}{M_1}\cos\left(\theta_1-\frac{2Ey}{A}\right)+\frac{\sqrt{M_1^2-E^2}}{M_1}\sin\left(\theta_1-\frac{2Ey}{A}\right)\right), \quad (45)$$

$$S_{yQD}(y) = \frac{2}{G}\left(-\frac{E}{M_1}\sin\left(\theta_1-\frac{2Ey}{A}\right)-\frac{\sqrt{M_1^2-E^2}}{M_1}\cos\left(\theta_1-\frac{2Ey}{A}\right)\right). \quad (46)$$

The expressions for the spin density inside the barriers have the form similar to (45), (46) except the additional factors describing the exponential decay into the barriers with the inverse lengths $\gamma_{1,2}$, respectively. Let us mention the properties of the spin density for the central level $E_0=0$ existing for antiparallel barrier configurations $\theta_2-\theta_1=\pi$. By setting $\theta_1=0$ from (45) and (46) we obtain that $S_{xQD}(y)\equiv 0$ and $S_{yQD}(y)=-2/G$, i.e. the spin density is uniformly aligned along the $Oy$ axis across the whole edge structure. This can be illustrated by the plots of charge density $|\psi(y)|^2$ and spin density shown in Figure 9 for the QW width $L=40$ nm. From Figure 9 one can conclude that both charge and spin density decay into the barriers on the scale of $1/\gamma_{1,2}$ which for the presently adopted parameters have similar values compared to the QD width. This result justifies the approximation of semi-infinite barriers in the Hamiltonian (30). For their width $b$ only the condition $b\gg 1/\gamma_{1,2}$ should be fulfilled which is easily accessible for the range of $b>100$ nm in the present model.

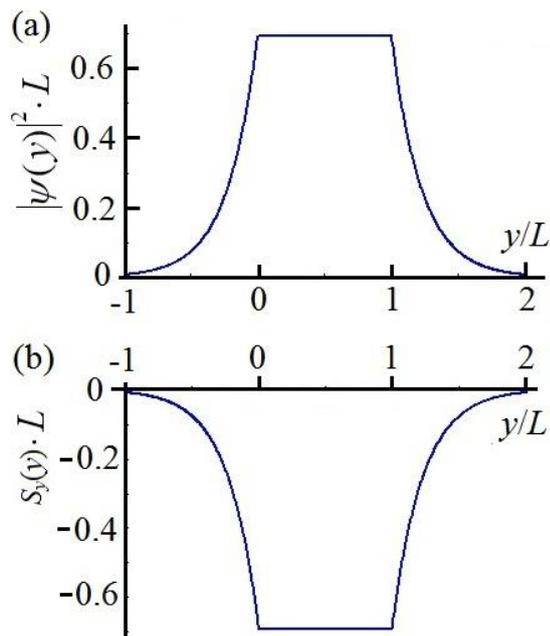

**Figure 9.** Spatial distributions of (a) charge density $|\psi(y)|^2$ and (b) non-zero spin density $S_y(y)$ in the QD area $0<y<L$ and in adjacent barrier regions shown for the central level $E_0=0$, existing for antiparallel barrier magnetizations $\theta_2-\theta_1=\pi$ in the Hamiltonian (30).

## 6. Conclusions

We have derived a model of bound state formation at the edge of 2D topological insulator by considering the interactions of finite size magnetic barriers with the topologically protected 1D edge states in a HgTe/CdTe quantum well. We have found that in such structure a 1D quantum dot is formed at the TI edge with variable number of bound states depending on barrier parameters. The spatial profile of exchange interaction between the edge states and barriers was obtained from the interaction with single impurity magnetic moment and is generalized for the barrier bulk structure formed by ensemble of impurities. The resulting Hamiltonian was studied as a function of barrier parameters including their strength and orientation of the magnetic moments. It was shown that for parallel magnetization of two barriers there are always at least two discrete levels being formed for any strength of the barrier while for antiparallel magnetization at least a single bound state is formed.

Our results indicate some possible experimental methods for the observation of the discrete level structure formed in the proposed QD. This can be performed, for example, by means of optical spectroscopy methods where the pronounced peaks of absorption should be observed on the frequencies corresponding to the transitions between the discrete levels. The typical interlevel distance in our model is comparable to the barrier height $M$ defined by the exchange coupling magnitude and is in the 1…10 meV range corresponding to the THz absorption range which is of big interest for nanostructure optics. For sufficiently high exchange coupling with the magnetic barriers when $M \sim 10$ meV the interlevel distance exceeds the thermal energy even for the liquid nitrogen environment which indicates the stable resolution of the predicted level structure at this widely-accessible thermal conditions. Such two- or three-level systems may serve as potential candidates for building the qubit or qutrit structures [41] based on QD formed in TI.

## Acknowledgements

The authors are grateful to M. M. Glazov, I. S. Burmistrov, S. A. Tarasenko, V. Ya. Aleshkin for very valuable discussions. The work is supported by the Ministry of Higher Education and Science of Russian Federation through the State Assignment No 0729-2020-0058 and by the President of Russian Federation grant for young scientists MK-3046.2022.1.2.